\documentclass[twocolumn,tighten]{aastex62}

\usepackage{blindtext}
\usepackage{graphicx}
\usepackage[colorinlistoftodos]{todonotes}
\usepackage{lineno}

\shortauthors{Madrid et al.}

\begin{document}


\title{Optical study of PKS~B1322-110, the intra-hour variable radio source}


\author{Juan P.\ Madrid}
\affiliation{Departmento de F\'isica y Astronom\'ia, La Universidad de Tejas de el Valle del R\'io Grande, Brownsville, TX 78520, USA}

\author{Artem V.\ Tuntsov}
\affiliation{Manly Astrophysics, 15/41-42 East Esplanade, Manly, NSW 2095, Australia}

\author{Mischa Schirmer}
\affiliation{Max-Planck Institut f\"ur Astronomie, K\"onigstuhl 17, D-69117 Heidelberg, Germany}

\author{Mark A. Walker}
\affiliation{Manly Astrophysics, 15/41-42 East Esplanade, Manly, NSW 2095, Australia}

\author{Carlos J.\ Donzelli}
\affiliation{Instituto de Astronom\'ia Te\'orica y Experimental IATE, CONICET - 
Observatorio Astron\'omico, Universidad Nacional de C\'ordoba, Laprida 854, X5000BGR, C\'ordoba, Argentina}

\author{Keith W. Bannister}
\affiliation{CSIRO, Astronomy and Space Science, PO BOX 76, Epping, NSW 1710, Australia}

\author{Hayley E.\ Bignall}
\affiliation{CSIRO, Astronomy and Space Science, 26 Dick Perry Avenue, Kensington, WA 6151, Australia}

\author{Jamie Stevens}
\affiliation{CSIRO, Astronomy and Space Science, Paul Wild Observatory, 1828 Yarrie Lake Road, Narrabri, NSW 2390, Australia}

\author{Cormac Reynolds}
\affiliation{CSIRO, Astronomy and Space Science, 26 Dick Perry Avenue, Kensington, WA 6151, Australia}

\author{Simon Johnston}
\affiliation{CSIRO, Astronomy and Space Science, PO BOX 76, Epping, NSW 1710, Australia}

\correspondingauthor{Juan P.\ Madrid }
\email{jmadrid@astro.swin.edu.au}

                         
\begin{abstract}

Observations with the Australia Telescope Compact Array revealed intra-hour
variations in the radio source PKS~B1322-110 \citep{bignall2019}. As part of an optical follow-up, 
we obtained Gemini H$\alpha$ and H$\alpha$ continuum (H$\alpha$C) images of the PKS~B1322-110 field. 
A robust 19~$\sigma$ detection of PKS~B1322-110 in the H$\alpha-$H$\alpha$C image prompted us to obtain 
the first optical spectrum of PKS~B1322-110. 
With the Gemini spectrum we determine that  PKS~B1322-110 is a flat-spectrum radio quasar at a redshift 
of $z=3.007\pm 0.002$. 
The apparent flux detected in the H$\alpha$ filter is likely
to originate from He\,{\sc ii} emission redshifted precisely on the Galactic H$\alpha$ 
narrow-band filter. We set upper limits on the emission measure of the Galactic plasma, for 
various possible cloud geometries.

\end{abstract}



\section{Introduction}

Intra-day variability (IDV) of radio quasars was recognized 30 years ago \citep{heeschen1987},
and became a topic of intense interest a decade later with the discovery of intra-hour variations
(IHVs) in PKS~0405-385 by \citet{kedziora1997}. The combination of large amplitude and short
timescale seen in that source strongly favoured scintillation as the cause, albeit manifesting
in a novel form. Following the discovery of additional IHV sources \citep[J1819$+$3845 and PKS1257$-$326;][]{dennettthorpe2000,bignall2003},
IHV was shown to be scintillation due to strongly scattering plasma clouds in the solar neighborhood
\citep{dennettthorpe2002, dennettthorpe2003, bignall2003,bignall2006}.


\begin{deluxetable*}{lcccc}
\tablecaption{Observation Log \label{obstable}} 
\tablehead{
\colhead{Program ID} & \colhead{Date} & \colhead{Filter} & \colhead{Grating} & \colhead{Exposure Time (s)}
}
\startdata
GS-2017A-Q-96        &  2017 Jun 28  & H$\alpha$ & None  &  8500 (10$\times$850)\\
GS-2017A-Q-96        &  2017 Jul 01  & H$\alpha$C & None  &  4250 (5$\times$850)\\
GS-2018A-FT-106      &  2018 Apr 09 & None      &  B600 &  2400 (3$\times$800)\\
GS-2018A-FT-106      &  2018 Apr 13 & None      &  R831 &   800 (1$\times$800)\\
GS-2018A-FT-106      &  2018 Apr 20 & None      &  R831 &  2400 (3$\times$800)\\
 \enddata

\end{deluxetable*}


The rarity of the IHV phenomenon was thus explained by the small probability for a 
line of sight to intersect such a cloud, and the more common IDV phenomenon is understood as being 
due to similar, but more distant plasma clouds --- with the smaller amplitude and longer timescale 
of IDV both attributable to the smoothing effect of the radio source size. These studies brought focus 
onto the clouds of dense plasma that are responsible for scattering the radio waves: What is their nature? 
In what physical context do they arise? And how are they energized?

Although it now seems likely that the plasma responsible for IDV and IHV is circumstellar, the material is located at
such large distances from the host stars -- with impact parameters of order 1 parsec -- that stellar winds cannot
explain the observed levels of radio-wave scattering \citep{walker2017}. Thus the nature of the plasma clouds remains
a mystery.

To date almost all of the information we have about the scattering plasma has come from studies of radio-wave
propagation --- either studies of IHV/IDV, or else studies of radio pulsars that seem to be revealing the same 
population of plasma clouds \citep{stinebring2001, cordes2006, walker2004, walker2008, brisken2010, tuntsov2013}. 
Even during drastic changes in radio flux density, optical observations show constant flux in those sources undergoing 
IHV/IDVs \citep[e.g.][]{bannister2016}.

If we could observe the intervening plasma directly, via its own thermal emission, then we could expect some immediate 
insights to follow from, e.g., density and temperature diagnostics in the emission line ratios, and from the size 
and morphology of the image of the cloud in any spectral line. These are attractive possibilities, but they 
present a substantial observational challenge for the following reasons. 

First, the plasma clouds appear to be of small 
spatial extent, with sizes of some tens of AU suggested by the observed transience of the IHV in J1819$+$3845 
\citep{debruyn2015}. Second, although the plasma is dense compared to the diffuse ISM, the available 
estimates of $n_e\sim10-30\,{\rm cm^{-3}}$ \citep{rickett2011,tuntsov2013} correspond to emission measures 
$n_e^2L\la0.1\,{\rm pc\,cm^{-6}}$, if $L\sim30\,{\rm AU}$ for example, and imply H$\alpha$ 
intensities $\la0.2\,{\rm R}$ \citep{reynolds2004}. \footnote{We measure line intensities in Rayleigh: 
$1\,{\rm R}~=~10^6/4\pi\,{\rm photons\,cm^{-2} \,s^{-1}\,sr^{-1}}$.}  

The expected emission measures are at or below the surface brightness limit that was achieved by the Wisconsin 
H-Alpha Mapper (WHAM) survey of H$\alpha$ emission \citep{reynolds2004}. WHAM data have relatively low angular
resolution (approximately $1^\circ$), whereas we require arcsecond resolution. 

Here we report our attempt to detect H$\alpha$ emission from the plasma responsible for the new IHV source 
PKS~B1322$-$110 using Gemini. Australia Telescope Compact Array observations recently demonstrated that PKS~B1322$-$110 
shows intra-day variability in its flux density between 4.3 and 11 GHz  \citep{bignall2019}.
PKS~B1322$-$110 is located at an angular distance of $\sim 8 \arcmin$ from Spica ($\alpha$ Virginis), the 16th brightest star in the sky.

\section{Gemini Observations and Data Reduction}
\subsection{Imaging}

Observations of PKS~B1322-110 were carried out with the Gemini Multiobject 
Spectrograph (GMOS) in imaging and spectroscopic mode from Gemini South. 


\begin{figure*}
        \centering
        \begin{tabular}{ccc} 
                \includegraphics[scale=0.215]{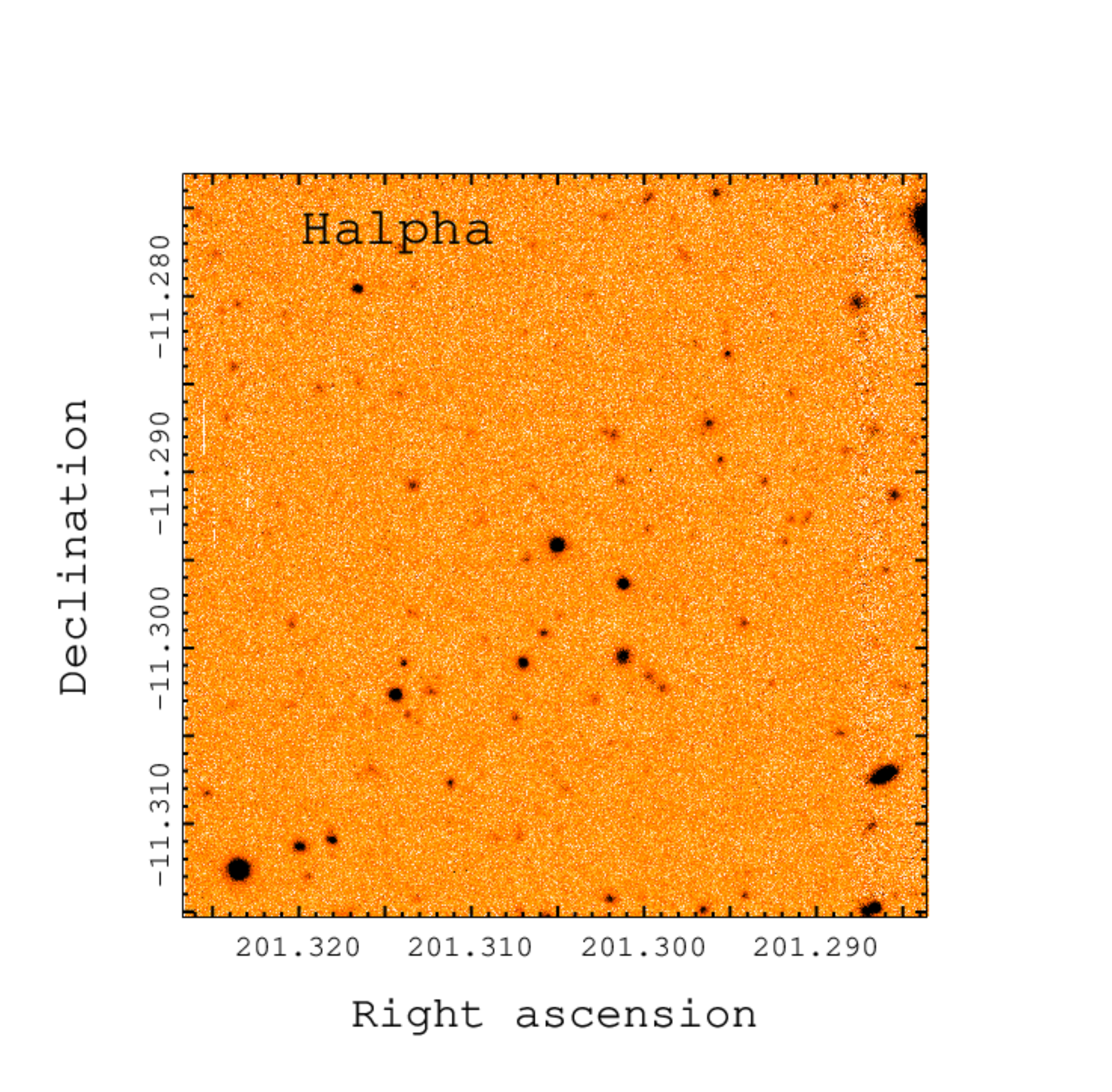} &  \includegraphics[scale=0.215]{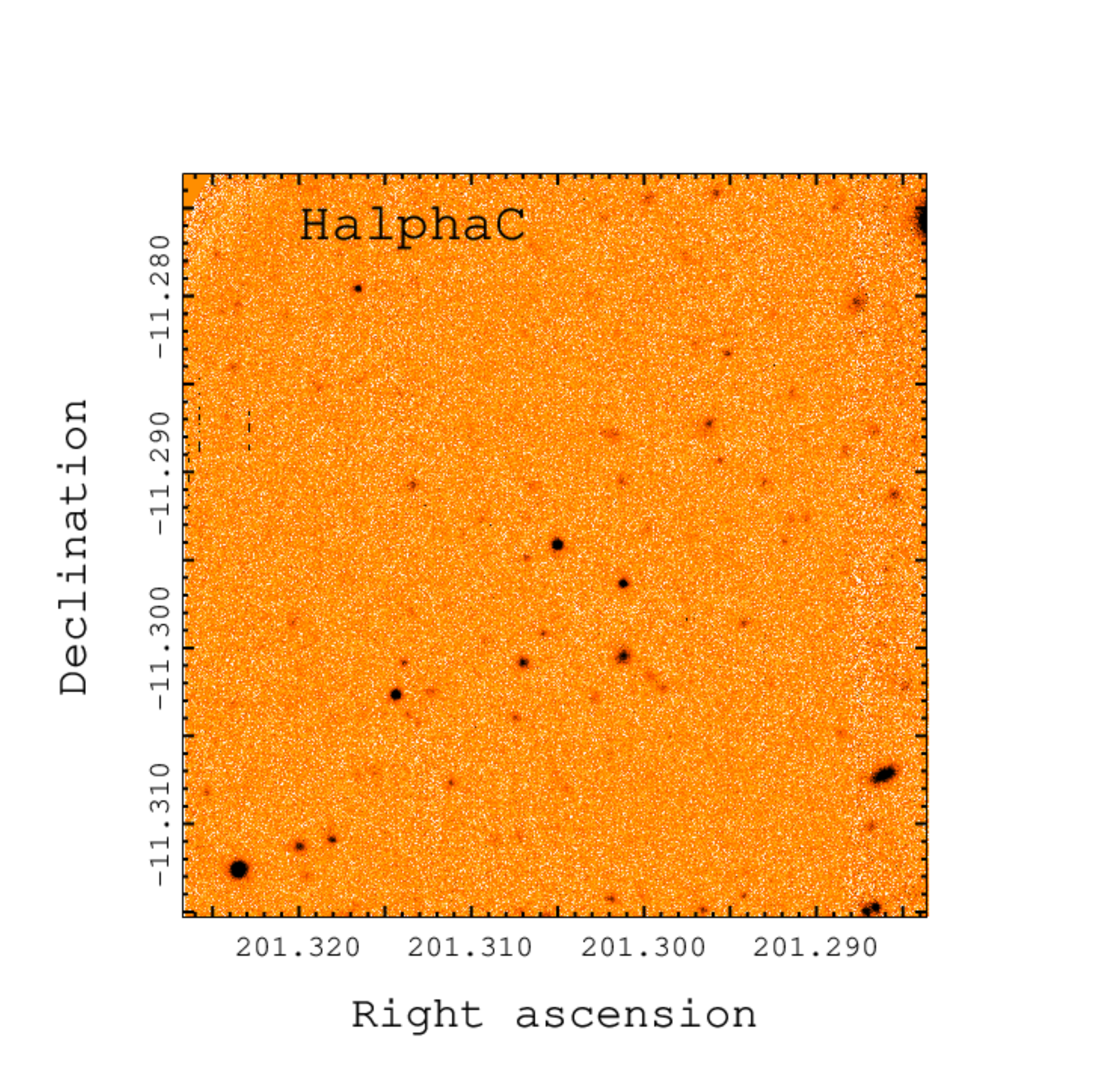} &  \includegraphics[scale=0.215]{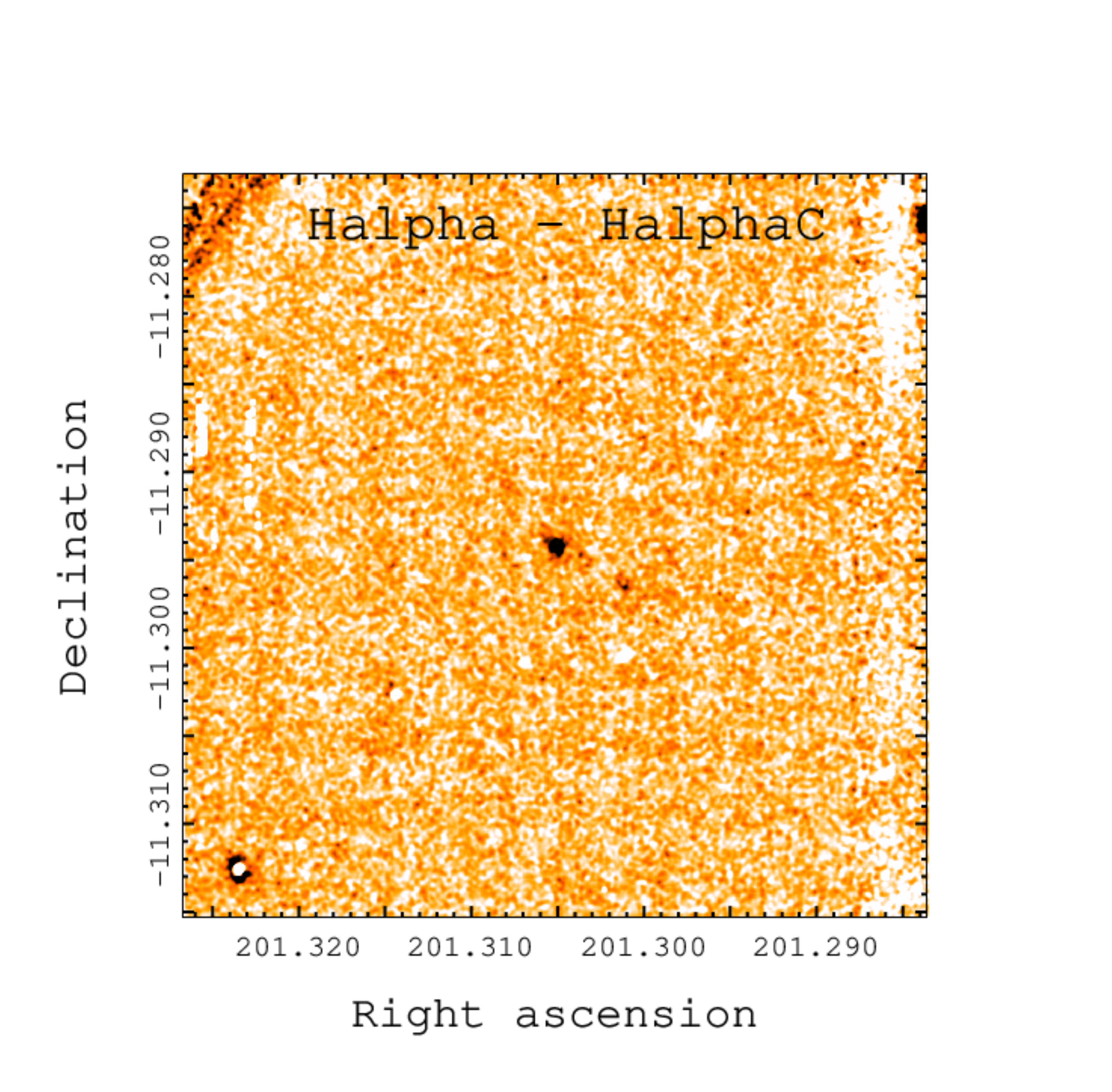}\\            
        \end{tabular}
    \caption{Gemini H$\alpha$ (\textit{left}) and H$\alpha$C (\textit{center}) images. 
    The subtraction of  the H$\alpha$ continuum from the  H$\alpha$ image (H$\alpha-$H$\alpha$C) is shown in the \textit{right panel}. 
    The residual image (H$\alpha-$H$\alpha$C) has been smoothed with a Gaussian kernel of five pixels for display purposes.
    All images are centered on PKS B1322-110, and show an area of 150$\times$150 arcseconds. PKS B1322-110 
    is the only source that shows a clear excess of H$\alpha$ flux on the section of the residual image shown above.}
    \label{imageshalpha}
\end{figure*}


Using GMOS, deep H$\alpha$ and H$\alpha$ continuum (H$\alpha$C) images centered on 
PKS B1322-110 were obtained. Details are given in Table 1.

The H$\alpha$ filter has its maximum throughput at 6560~\AA, the rest-frame wavelength of H$\alpha$ 
emission. For the H$\alpha$ filter the wavelength interval is 6540--6610 \AA~(width at 
half-maximum). The H$\alpha$C filter is centered at 6620~\AA~with 
a transmission interval of 6590--6650 \AA.

The imaging data were obtained on UT 2017 May 28 (H$\alpha$, cirrus, 0.51
mag transmission variation peak-to-valley) and UT 2017 August 01 (H$\alpha$ continuum,
clear sky). On both nights the sky
conditions were dark (Gemini \textsc{SB50} sky brightness quantile).
The seeing in the coadded images is $1.14^{\prime\prime}$ (H$\alpha$) 
and $1.22^{\prime\prime}$ (H$\alpha$ continuum).

The data were reduced using the THELI data reduction pipeline \citep{erben2005,schirmer2013}.
Standard bias and flat-field correction were applied. The presence of
the bright star Spica ($\alpha$ Virginis, $V=0.98$ mag) just 8$^{\prime}$ 
north of our target required us to model and subtract the sky background of the 
individual dithered exposures. This was done following THELI's standard recipes for
data processing \citep{schirmer2013}.

A common astrometric solution was obtained for both filters using \textsc{Scamp} 
\citep{bertin2006}. \textsc{Scamp} computes astrometric and photometric solutions 
for any FITS image in an automated way. The individual exposures were registered against 
the Gaia DR1 astrometric reference catalog. After distortion correction, the images were 
registered within 1/15th of a pixel with respect to each other.
The H$\alpha$ and the H$\alpha$C images were processed simultaneously 
omitting their filter information. This allows \textsc{Scamp} to determine the mean relative
throughput difference of the two filters based on stellar magnitudes in the field. 
The ansatz for our method  is that the difference of the two filters should yield a null flux 
for stars. By following this procedure, the fluxes of the on-band (H$\alpha$) 
and off-band (H$\alpha$C) images were already calibrated, and deriving their difference 
image is a straightforward subtraction.\\

\subsection{Spectrophotometric calibration}

In the absence of observations of a spectrophotometric standard star, a flux calibration of
the narrowband images was performed as follows. First, we integrate the effective
H$\alpha$ and H$\alpha$ continuum transmission curves including detector quantum efficiency, 
and find that both narrowband filters transmit a factor 20.3 less than the GMOS $r$-band
filter. We upscaled the two coadded images by that factor, and compared the $r$-band aperture 
magnitudes in these images with the PanSTARRS-DR1 $r$ magnitudes of the same field. 


\subsection{Spectroscopy}

\begin{figure*}
\epsscale{0.9}
 \plotone{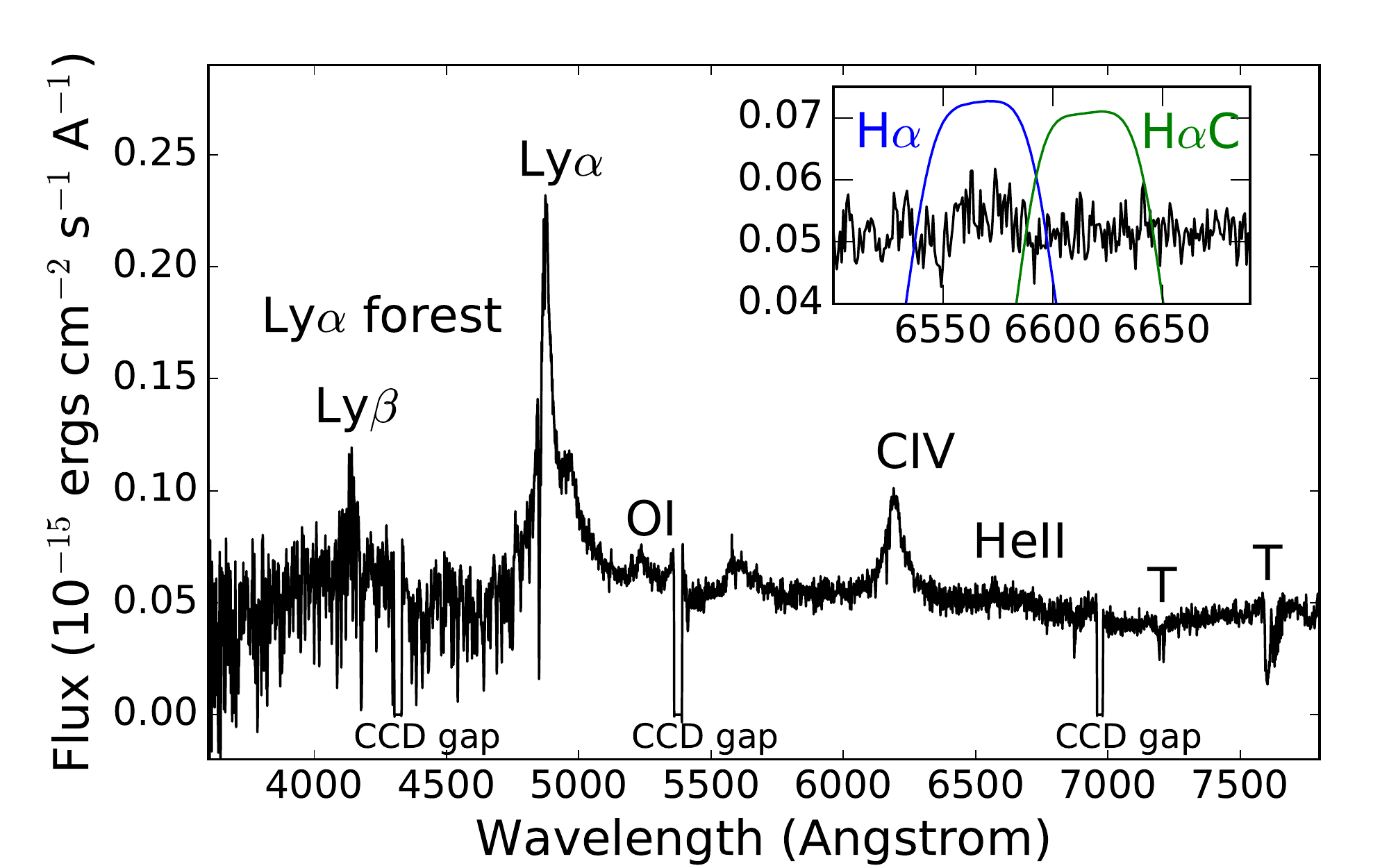}
 \caption{GMOS spectrum of PKS B1322-110. This spectrum shows the combined, calibrated 
 data obtained with both gratings, i.e.\ B600 and R831. The letter {\sc T} denotes prominent telluric
 absorption lines that are present, as expected,  at wavelengths larger than 6700 \AA. The inset zooms on the section of the spectrum at the wavelength of Galactic H$\alpha$.
 In this inset, the transmission curves for the H$\alpha$ (blue line) and H$\alpha$ continuum (green line)
 filters are overplotted. At the redshift of $z$=3.0 the He\,{\sc ii} line falls precisely in the passband
 of the Galactic H$\alpha$ filter. The spectrum is displayed at the observed wavelength.
 \label{gmosspectrum}}
\end{figure*}

GMOS spectra of PKS~B1322-110 were obtained with two gratings: B600 and R831. 
The B600 grating has a blaze wavelength of 4610\AA.  
Likewise, the R831 grating has a blaze wavelength of 7570\AA. This
observing configuration is adopted in order to have a wide spectral coverage with a
high signal-to-noise from $\sim$3500 to $\sim$8000 \AA, as shown in the final spectrum in 
Figure ~\ref{gmosspectrum}. 

Spectroscopic data were reduced by applying a sequence of standard {\sc pyraf} tasks 
available through the Gemini package. The data reduction procedure used here has been 
described in detail in \citet{madrid2013}. The spectra of the spectrophotometric 
standard star {\sc EG274} are used to flux calibrate the science spectra.

\section{Results}

The narrowband images, H$\alpha$ and H$\alpha$ continuum, are presented in 
 Figure \ref{imageshalpha}. The H$\alpha$ narrowband filter is centered on the rest-frame 
wavelength of H$\alpha$ (6560 \AA). The H$\alpha$ continuum filter is adjacent to 
the H$\alpha$ filter and it is designed to allow for the removal of the underlying 
continuum. By obtaining images with these two filters and computing their
difference (H$\alpha$~$-$~H$\alpha$C) one can detect sources with excess H$\alpha$ 
emission.

Both the H$\alpha$ and H$\alpha$ continuum images show dozens of sources in the field,
most of these sources are galaxies or stars; see Fig.~\ref{imageshalpha}. On the other
hand, the difference image (H$\alpha-$H$\alpha$C) has only one clear source on the 
entire central part of the frame, also shown in Fig.~\ref{imageshalpha}.

The clear detection of excess flux in the H$\alpha$ filter (19~$\sigma$) coincident with PKS~B1322-110 was compelling 
and prompted  us to request additional time on Gemini South in order to obtain 
a spectrum of PKS~B1322-110. Also, until now, the nature and redshift of PKS~B1322-110
have remained unknown.

The calibrated GMOS spectrum of PKS~B1322-110 obtained during the spectroscopic campaign
is presented in Fig.~\ref{gmosspectrum}. This spectrum shows a prominent Ly$\alpha$ line and the associated
 Ly$\alpha$ forest. The Ly$\beta$ and C\,{\sc iv} emission lines are also prominent. Using the above emission lines, jointly 
with other prominent lines like O\,{\sc i}, and N\,{\sc v} we determine that the redshift of PKS~B1322-110 is  $z=3.007 \pm 0.002$. 
The optical spectrum of PKS~B1322-110 is characteristic of a quasar; see for instance the composite 
quasar spectra templates from the SDSS \citep{vandenberk2001}. Considering its radio properties \citep{griffith1994}
PKS~B1322-110 can be classified as a flat spectrum radio quasar (FSRQ).

Given that the redshift of PKS~B1322-110 is now determined to be $z$=3.007 the presence
of an excess flux at 6560~\AA, the rest-frame wavelength of H$\alpha$, is likely due to 
the redshifted emission of the He\,{\sc ii} line. He\,{\sc ii} is an emission
line with a rest-frame wavelength of 1640~\AA~that is commonly found on quasar
spectra \citep[e.g.][]{wilkes1986, jakobsen2003}. The observed emission of the He\,{\sc ii}
line at the redshift of PKS~B1322-110  is $\lambda_{obs}~=~1640~\times~(1+z)=6560$ \AA, falling 
exactly on the passband of the Galactic H$\alpha$ narrowband filter.

At a redshift of $z$=3.007, PKS~B1322-110 is at the high end of the redshift distribution 
of scintillating radio sources. Indeed, \citet{lovell2008} found that radio sources that 
show interstellar scintillation drop steeply beyond redshift $\sim$2.


\section{Line emission constraints}


\subsection{An upper limit for the H$\alpha$ flux}

The Gemini spectrum of PKS~1322$-$110, displayed in Figures~2 and~3, lacks any conspicuous 
narrow line emission at the wavelength of Galactic H$\alpha$, where emission from the plasma 
responsible for radio scintillations might be expected. Using the Gemini spectrum we derive 
an upper limit for the H$\alpha$ flux hypothetically emanating from the intervening plasma 
that might be blended with the active galactic nucleus (AGN). These upper limits are derived by measuring the variation 
in the spectrum from one resolution element to the next. The resolution element corresponds 
to three pixels, that is, $2.2\mathrm{\AA}$ or $102\,\mathrm{km}\,\mathrm{s}^{-1}$ at the 
H$\alpha$ wavelength (approximately 6563$\AA$ in air). The standard deviation of the 
spectrum is $\approx3.4\times10^{-18}\,\mathrm{erg}\,\mathrm{cm}^{-2}\,\mathrm{s}^{-1}\,\mathrm{\AA}^{-1}$. 
This value is measured within $10\,\mathrm{\AA}$ to either side of H$\alpha$ in the spectrum 
running-averaged to the resolution of $2.2\,\mathrm{\AA}$. Integrating over the line, 
we conclude that the $3\sigma$ upper limit to the H$\alpha$ flux of plasma that might be blended 
with the AGN is $2.3\times10^{-17}\,\mathrm{erg}\,\mathrm{cm}^{-2}\,\mathrm{s}^{-1}$.

We note that the value of the standard deviation given above is larger than our estimate 
of the noise in the spectrum, which is $\approx1.5\times10^{-18}\,\mathrm{erg}\,\mathrm{cm}^{-2}\,\mathrm{s}^{-1}\,\mathrm{\AA}^{-1}$. 
This difference presumably reflects intrinsic structure in the AGN spectrum. 

\subsection{Spatial extent of the intervening plasma}

The limit on the emission measure of the plasma depends on its assumed spatial extent. 
As the scintillations of PKS~B1322-110 have been sustained for at least 2 years 
\citep{bignall2019}, the proper motion of the scattering plasma sets a lower limit 
on the size of the cloud responsible for the scintillations. Assuming the plasma is 
comoving with Spica -- which is consistent with the measured annual cycle of PKS~B1322-110 
\citep{bignall2019} -- this lower limit is $\simeq100$ milliarcseconds (mas) in one direction on the sky. 

In the case of a  the plasma cloud that is at least $100\,\mathrm{mas}$ by $100\,\mathrm{mas}$ its 
brightness upper limit is $410\,\mathrm{R}$. To convert brightness to emission measure we assume 
Case B recombination at $T=10^4\,\mathrm{K}$, giving a conversion factor of $0.361\,\mathrm{R}\,\mathrm{cm}^6\,\mathrm{pc}^{-1}$ 
(Table~14.2 of \citealt{draine2011}). The emission measure upper limit is  then  $1.1\times10^3\,\mathrm{cm}^{-6}\,\mathrm{pc}$.

The foregoing calculation applies to a compact plasma cloud whose emission is blended with that of 
the background quasar. We searched for extended emission by modeling the long-slit data obtained 
by Gemini as a sum of a compact source (convolved with a PSF along the slit) and a sky background 
that is flat along the slit. We remove the model from the observed data and analyze the residual. 
No extended emission can be seen at the wavelength of H$\alpha$ down to the level 
$\sim1.9\times 10^{-17}\,\mathrm{erg}\,\mathrm{cm}^{-2}\,\mathrm{s}^{-1}\,\mathrm{\AA}^{-1}\,\mathrm{arcsec}^{-2}$. 
For emission wider than the slit this corresponds to the intensity density $3\sigma$ limit 
of $10\,\mathrm{R}\,\mathrm{\AA}^{-1}$; integrating over the spectral resolution element, 
one obtains the intensity limit of $22\,\mathrm{R}$ equivalent to $\mathrm{EM}\simeq62\,\mathrm{cm}^{-6}\,\mathrm{pc}$ 
under the stated equilibrium conditions. This estimate assumes a unit surface filling 
fraction maintained on arcsecond spatial scales. 

A third scenario for the spatial extent of the intervening plasma is that it is uniform and extended 
over the entire length of the slit. In this case, any plasma emission would have been 
subtracted with the sky modeling and could not be seen in the residuals. The sky spectrum does display 
a line at the wavelength of H$\alpha$, which integrates to $17\,\mathrm{R}$ above the estimated continuum 
level of $24\,\mathrm{R}\,\mathrm{\AA}^{-1}$. At the spectral resolution of $\sim2\,\mathrm{\AA}$ we cannot 
distinguish Galactic H$\alpha$ emission from that of the night sky and therefore can only set an upper 
limit on the intensity of the intervening plasma emission at $17\,\mathrm{R}$, which corresponds to the 
emission measure of $47\,\mathrm{cm}^{-6}\,\mathrm{pc}$ -- again assuming a unit surface filling fraction, 
this time sustained over a scale of a few arcminutes.


\begin{figure}
	\includegraphics[width=\columnwidth]{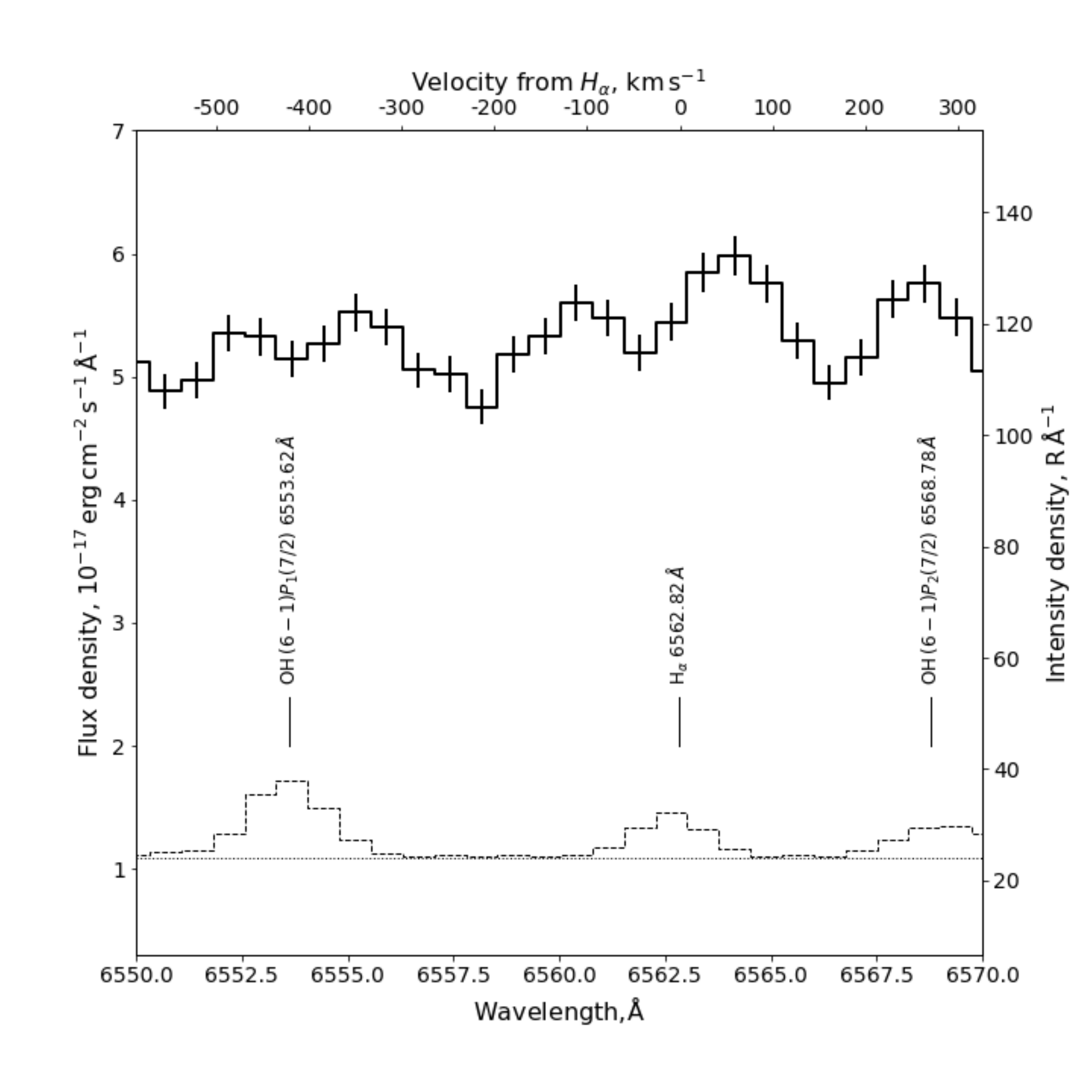}
    \caption{Zoom-in of the AGN (solid) and sky background (dashed) spectra of PKS~1322$-$110 within $20\mathrm{\AA}$ adjacent to 
    the Galactic H$\alpha$ position; the horizontal dotted line shows our estimate of the sky continuum level. Line identifications are given 
    for the night sky background, based on \citet{osterbrock1996} and \citet{hanuschik2003}.}
    \label{figure:spectralzoom}
\end{figure}

\section{Conclusion}
Gemini narrowband imaging resulted in an apparent H$\alpha$ detection toward the 
fast scintillator PKS B1322-110, and we therefore undertook follow-up spectroscopy with Gemini.
The spectroscopy demonstrated that Galactic H$\alpha$ emission was not responsible for the signal 
that we observed in our images. In part the signal we observed in our imaging data is due to He\,{\sc ii} 
emission in the quasar redshifted to the wavelength of Galactic H$\alpha$. 
We have determined limits on the possible H$\alpha$ surface brightness of the intervening Galactic 
plasma cloud, depending on its size. However, even in the case of a very
extended cloud the corresponding limits on emission measure are well above the values expected 
from modeling the scintillations of other radio sources.\\

\acknowledgments

We thank Nathan Pope (CSIRO) for his help with computing resources.
Based  on observations  obtained at  the Gemini  Observatory  which is 
operated by AURA under a cooperative  agreement with  the NSF  on 
behalf  of the Gemini partnership: the National Science  Foundation (United States), 
the   National  Research  Council   (Canada),  CONICYT   (Chile), Minist\'{e}rio 
da Ci\^{e}ncia, Tecnologia e  Inova\c{c}\~{a}o (Brazil) and Ministerio de 
Ciencia, Tecnolog\'{i}a e Innovaci\'{o}n Productiva (Argentina).






\end{document}